\documentstyle[emulateapj,psfig]{article}

\begin{document}

\lefthead{Brotherton et al.}
\righthead{Post-Starburst Quasar}

\title{A Spectacular Post-Starburst Quasar}
\author{M. S. Brotherton\altaffilmark{1}, Wil van Breugel\altaffilmark{1}, S. 
A. Stanford\altaffilmark{1}, R. J. Smith\altaffilmark{2}, B. J. 
Boyle\altaffilmark{3}, Lance Miller\altaffilmark{4}, T. Shanks\altaffilmark{5}, S. M. Croom\altaffilmark{6}, \&  Alexei V. Filippenko\altaffilmark{7}}

\altaffiltext{1}{Institute of Geophysics and Planetary Physics, Lawrence 
Livermore National Laboratory, 7000 East Avenue, P.O. Box 808, L413, Livermore,
CA 94550; mbrother@igpp.llnl.gov, wil@igpp.llnl.gov, adam@igpp.llnl.gov}
\altaffiltext{2}{Research School of Astronomy and Astrophysics,
Australian National University, Private bag, Weston Creek P.O., ACT 2611,
Australia; rsmith@mso.anu.edu.au}
\altaffiltext{3}{Anglo-Australian Telescope, PO Box 296, Epping, NSW 2121, 
Australia; director@aaoepp2.aao.gov.au}
\altaffiltext{4}{Department of Physics, Oxford University, Keble Road, Oxford, 
OX1 3RH, UK; l.miller1@physics.ox.ac.uk}
\altaffiltext{5}{Department of Physics, University of Durham, South Road, 
Durham, DH1 3LE, UK; Tom.Shanks@durham.ac.uk}
\altaffiltext{6}{Imperial College of Science, Technology and Medicine,
Blackett Laboratory, Prince Consort Road, London, SW7 2BZ, UK; s.croom@ic.ac.uk}
\altaffiltext{7}{Department of Astronomy, University of California, Berkeley,
CA 94720-3411; alex@astro.berkeley.edu}

\begin{abstract}

We report the discovery of a spectacular ``post-starburst quasar'' 
UN J1025$-$0040 ($B=19$; $z=0.634$).  The optical spectrum is a chimera, 
displaying the broad Mg II $\lambda$2800 emission line and strong blue 
continuum characteristic of quasars, but is dominated in the red by a large 
Balmer jump and prominent high-order Balmer absorption lines indicative of 
a substantial young stellar population at similar redshift.
Stellar synthesis population models show that the stellar component is
consistent with a 400 Myr old instantaneous starburst with a mass of
$\la$10$^{11}$ $M_{\sun}$.  A deep, $K_s$-band image taken in
$\sim0.5 \arcsec$ seeing shows a point source surrounded by asymmetric
extended fuzz.  Approximately 70\% of the light is unresolved, the majority of 
which is expected to be emitted by the starburst.
While starbursts and galaxy interactions have
been previously associated with quasars, no quasar ever before has been 
seen with such an extremely luminous young stellar population.

\end{abstract}

\section{Introduction}

Is there a connection between starbursts and quasar activity?  
There is circumstantial evidence to suggest so.
3C 48 is surrounded by nebulosity that shows 
the high-order Balmer absorption lines characteristic of A-type stars 
(Boroson \& Oke 1984; Stockton \& Ridgeway 1991).  PG 1700+518 shows a nearby 
starburst ring (Hines et al. 1999) with the spectrum of a
$10^8$-year-old starburst (Stockton, Canalizo, \& Close 1998).
Near-IR and CO mapping reveals a massive ($\sim10^{10} M_{\odot}$) 
circumnuclear starburst ring in I Zw 1 (Schinnerer et al. 1998).  
The binary quasar member FIRST J164311.3+315618B shows a starburst 
host galaxy spectrum (Brotherton et al. 1999).

In addition to these individual objects, {\em samples} of active galactic 
nuclei (AGNs) show evidence of starbursts.
Images of quasars taken with the {\em Hubble Space Telescope} show 
``chains of emission nebulae'' and ``near-nuclear emission knots''
(e.g., Bahcall et al. 1997).  Seyfert 2 and radio galaxies 
have significant populations of $\sim$100 Myr old stars 
(e.g., Schmitt, Storchi-Bergmann, \& Cid Fernandes 1999).
Half of the ultraluminous infrared galaxies (ULIRGs) 
contain simultaneously an AGN and recent (10-100 Myr) starburst activity 
in a 1-2 kpc circumnuclear ring (Genzel et al. 1998).

The advent of $IRAS$\ provided evidence for an evolutionary link between
starbursts and AGNs. The ULIRGs ($L_{IR}$ $> 10^{12} L_{\odot}$) are strongly 
interacting merger systems with copious molecular gas 
[($0.5-2)\times10^{10} M_{\odot}$] and dust heated by both
starburst and AGN power sources.  The ULIRG space density is sufficient to
form the quasar parent population.  These facts led Sanders et al. (1988)
to hypothesize that ULIRGs represent the initial dust-enshrouded stage of a
quasar.  Supporting this hypothesis is the similarity in the 
evolution of the quasar luminosity density and the star formation rate 
(e.g., Boyle \& Terlevich 1998; Percival \& Miller 1999).  Another clue is that 
supermassive black holes appear ubiquitously in local massive galaxies, 
which may be out-of-fuel quasars (e.g., Magorrian et al. 1998).
AGN activity may therefore reflect a fundamental stage of galaxy evolution.

We report here the discovery of a post-starburst quasar.
The extreme properties of this system may help shed light on the elusive 
AGN-starburst connection.
We adopt $H_0 = 75$ km s$^{-1}$ Mpc$^{-1}$ and $q_0$ = 0.

\section{Observations}
\subsection{Selection and Spectroscopy}
We identified UN J1025$-$0040 as a quasar candidate in the UVX-NVSS (UN) 
catalog: entries include candidates from the ultraviolet excess 
(UVX) AAT-2dF quasar survey (Smith et al. 1996) within 10$\arcsec$
of a radio source from the NRAO VLA Sky Survey (NVSS; Condon et al. 1998).
UN J1025$-$0040 was identified as a stellar optical source, with
$B_j=19.34, U-B_j = -0.28$, and $B_j-R=0.93$. The optical position is 
$\alpha$=10h25m00.78s, $\delta$=$-$00$\arcdeg$40$^\prime$44$\farcs$9 (J2000), 
6$\arcsec$ from an unresolved NVSS source with a 20 cm flux density of 
9.2$\pm$0.6 mJy.  A new map from the FIRST survey (Becker, White, \& Helfand 
1995) shows a 20 cm source with $\sim$ 3 mJy and a separation of less than an 
arcsecond from the optical counterpart (R. Becker, priv. comm.).  
The difference in positions and flux densities of the two radio surveys are
consistent with the presence of extended emission ``resolved out'' at high 
resolution.

We obtained 4-minute Keck ``snapshot'' spectra of UN J1025$-$0040 and 
other quasar candidates from the equatorial (+2.5$\arcdeg$ to $-$2.5$\arcdeg$) 
UVX-NVSS catalog with $18.25 < B_j \leq 20.00$ and right ascensions from 
10--13 hours (see Brotherton et al. 1998).
Observations were conducted with the Low Resolution Imaging Spectrometer (LRIS;
Oke et al. 1995) on 1998 February 9 (UT) with the slit oriented north-south.
The 300 line mm$^{-1}$ grating blazed at 5000 \AA\ with a
1$\arcsec$ slit gave an effective resolution $\leq$ 10 \AA\ 
covering 4000 \AA\ to 9000 \AA. 
We employed standard data reduction techniques within the NOAO IRAF package.

We also obtained spectropolarimetry using LRIS at the Keck II telescope on
1998 March 6 (UT).  The 60 minute observation was broken into four 15 minute 
exposures, one for each waveplate position
(0$\arcdeg$, 45$\arcdeg$, 22.5$\arcdeg$, 67.5$\arcdeg$).
The 600 line mm$^{-1}$ grating blazed at 5000 \AA\ with a 1$\arcsec$ slit 
gave an effective resolution $\leq$ 5 \AA\ covering 4300 \AA\ to 6800 \AA.
We used standard reduction techniques inside the IRAF NOAO package and
the procedures of Miller, Robinson, \& Goodrich (1988) for calculating Stokes 
parameters and errors.  Binning the data (in $q$\ and $u$) from 4500-6500 \AA\ 
yields a polarization of $p=1.28\pm0.05\%, \theta=62.5\arcdeg\pm1.2\arcdeg$.  
We have sufficient S/N ratio to distinguish between ``blue'' and ``red'' 
polarizations: binning from 4400-5400 \AA\ (below the redshifted Balmer jump) 
yields $p=1.0\pm0.1\%, \theta=61.3\arcdeg\pm2.2\arcdeg$, while binning from 
6200-6700 \AA\ (above the redshifted Balmer jump)  yields
$p=1.6\pm0.1\%, \theta=66.1\arcdeg\pm1.6\arcdeg$.

The Galactic reddening in the direction of UN J1025-0040
is $E(B-V) = 0.057$ mag (Schlegel, Finkbeiner, \& Davis 1998).  
The {\em maximum} expected interstellar polarization is $9 \times E(B-V)$\%
(Serkowski, Mathewson, \& Ford 1975), or 0.5\%.  The
polarization in UN J1025-0040 is therefore intrinsic.

Figure 1 (top) shows the total-light spectrum of UN J1025$-$0040.

\subsection{Near-Infrared Imaging}

We imaged UN J1025$-$0040 in $K_s$ using NIRC (Matthews \& Soifer 1994) 
at the Keck I telescope on 18 April 1998 when 
conditions were clear with $\sim$0\farcs5 seeing. 
The total on-source exposure time was 1920~s.  
After bias subtraction, linearization, and
flatfielding (using sky flats), the data were sky--subtracted,
registered, and summed using DIMSUM\altaffilmark{1}.
Subsequent photometry was flux calibrated via
observations of UKIRT faint standards (Casali \& Hawarden 1992), which, 
after a suitable transformation, yields magnitudes on the Caltech system
(Elias et al. 1982).  
The image (Fig. 2) shows a strong unresolved source surrounded by
asymmetric extended fuzz with a feature suggestive of a dust lane, 
and a companion ($K_s = 20.7$, $4\farcs5$ to the SSW).

\altaffiltext{1}{DIMSUM is the Deep Infrared Mosaicing Software package,
developed by P.\ Eisenhardt, M.\ Dickinson, A.\ Stanford, and J.\ Ward,
which is available as a contributed package in IRAF.}

\section{Spectral Analysis: Starburst and Quasar}

We explored the stellar synthesis population models of Bruzual \& Charlot (1999)
in order to interpret the ``A-star'' spectrum of UN J1025$-$0040.
The size of the Balmer jump, even diluted by a quasar continuum,
argues for a population at least several times $10^8$ years old.  The fact that
there is a Balmer jump rather than a 4000 \AA\ break suggests that the age
is less than $\sim$0.5 Gyr.  A solar metallicity, Salpeter initial mass function
(0.1 to 125 $M_{\sun}$), 400 Myr old instantaneous starburst (ISB) model
reproduces the observed features rather well (Fig. 1).  
The mass of the starburst, as determined by the model mass and its scaling
to the observed spectrum, is large: 7$\times$10$^{10}$ $M_{\sun}$.

Although this very simple model matches the observed features rather well, 
there are additional complexities to keep in mind. If the metallicity is much 
higher than solar, a younger ($\ga$350 Myr) stellar population can reproduce 
the spectrum, while a lower metallicity allows for an older population 
($\la$ 450 Myr).  The modeling at these epochs is not very sensitive to
the choice of initial mass function.  A second population, such as that
of the old underlying host galaxy, probably contributes at some level,
but is not yet required in the modeling. Also not yet required is the 
invocation of dust reddening, although it is possible that some is present 
in one or more components.  These effects can raise or lower the mass 
estimate by factors of a few.

While the Balmer absorption lines and [O III] $\lambda\lambda$4959,5007 have 
the same redshift of $z=0.634$, the broad (FWHM = 5500$\pm$500 km s$^{-1}$)
Mg II $\lambda$2800 is at $z=0.627$.  After subtracting off
the ISB model from the low-resolution spectrum (bottom panel Fig. 1), 
broad H$\beta$ is revealed, also with a peak at $z=0.627$.  Low-ionization 
broad lines blueshifted by $\sim$1000 km s$^{-1}$ are not unprecedented 
(e.g., Marziani et al. 1996). 
Also worth noting is that the line intensity ratio 
Mg II $\lambda$2800/H$\beta$ $\approx$ 5,
while values of 0.5-2.5 are more typical for quasars (Grandi \& Phillips 1979).
Otherwise the quasar spectrum appears typical, if slightly bluer than
average.

\section{Image Analysis: Point Source and ``Fuzz''}

We have performed a one-dimensional profile analysis of our image
in order to separate the unresolved nucleus from the surrounding fuzz.
We used the ellipse fitting procedure in IRAF STSDAS to generate a 
radial intensity profile of UN J1025-0040.  
An appropriate star with a similar count rate to UN J1025-0040 was used
to generate a PSF profile.  
The surface brightness profile limit is about 22 mag arcsec$^{-2}$.
We subtracted different amounts of this
PSF from the UN J1025-0040 profile, trying to find values for which the
residual would be linear with $r$\ (disk fit) or $r^{1/4}$ (De Vaucouleurs fit).
For both fits, 70\% of the 
normalized PSF is attributable to an unresolved nucleus (Fig. 3).
Given that the total magnitude of UN J1025-0040 is $K_s=16.74$, the nucleus 
has a magnitude of $K_s=17.1$, and the underlying fuzz has a magnitude of
$K_s=18.0$ (with an uncertainty of at least 0.2 mag).
The ISB model suggests that the ratio of starburst to quasar light in the
$K_s$-band is $\sim$ 3.  Since 70\% of the total $K_s$-band light is 
unresolved, a significant part of the starburst must also be unresolved.  
We see only one nucleus, so a portion of the starburst is within 
$\sim$ 2 $h^{-1}$ kpc (in projection) of the quasar.
The ISB model predicts $K_s$=17.0, consistent with the
unresolved component, but there must be some contribution from the quasar.
This implies that part of the starburst may be extended.

\section{Discussion}

Table 1 summarizes some observed and derived properties of UN J1025$-$0040.
The most straightforward interpretation of our results
is that UN J1025$-$0040 is an interacting galaxy harboring both a quasar
and a massive, $\sim$400 Myr old nuclear starburst.  As the first quasar found
in tandem with such an {\em extremely} luminous young stellar population
($M_B=-24.7$), the object's existence alone is of importance.  
We classify UN J1025$-$0040
as a ``post-starburst'' quasar in analogy with ``post-starburst''
or ``k + a'' galaxies (Dressler et al. 1999), which display prominent A-star 
populations, but no [O II] $\lambda$3727 emission, suggesting they are
without on-going star formation.

We do not know for certain that there is no current star formation in
UN J1025$-$0040.  The quasar could ionize O$^+$ to O$^{++}$ 
(as [O III] $\lambda$5007 is seen),
and the presence of young, hot stars could be masked by the quasar continuum.
We searched the $IRAS$\ database using ADDSCAN, and found no emission at 
60 $\micron$ at a level of 0.15 Jy (3$\sigma$).  The formalism of 
Sanders \& Mirabel (1996) gives L$_{FIR}$ $<$ 10$^{12.4}$ $L_{\sun}$. 
The bolometric luminosity of the integrated ISB model is $10^{11.6}$ 
$L_{\sun}$, consistent with the non-detection even if there is significant dust
present.  UN J1025-0040 sits above the star-forming track on the 
radio/far-infrared luminosity diagram (van Breugel 1999), 
where an AGN is expected to account for the large radio emission
(L$_{1.4GHz}$ = 10$^{32.2}$ ergs s$^{-1}$ Hz$^{-1}$).
Using the spectral energy distributions of Elvis et al. (1994)
to predict a bolometric luminosity for the quasar residual based on its
optical flux, we obtain $10^{11.6\pm0.5}$ $L_{\sun}$, the same as that of the
ISB model within the uncertainties. 

The dust content is uncertain, and will require additional infrared observations
to constrain.  Dust is a common feature of starbursts, and there is 
circumstantial evidence for its presence.  The polarization that rises to the 
red could result from the passage of light through aligned dust grains 
associated with the starburst, although other hypotheses are viable.  
Moreover, the dark streak in the fuzz to the
SW, while only marginally detected, may represent a dust lane.

The frequency of objects like UN J1025$-$0040 is unclear, as 
less extreme versions require detailed observations to recognize.
A post-starburst quasar could have been found in previous surveys and
misidentified because of low-quality spectra or limited wavelength coverage.
Because of its age ($\sim$400 Myr), the starburst could only have been 
brighter and bluer in the past.
Such extreme starbursts have not been identified in previous ultraviolet 
excess quasar surveys, suggesting that a younger version of UN J1025$-$0040 
might be hidden as a dust-enshrouded ULIRG.  If we rerun
the ISB model to an age of 55 Myr (old enough that an instantaneous starburst 
is likely still a good description of the event), the bolometric luminosity is 
$10^{12.4} L_{\sun}$, consistent with the infrared luminosity of an ULIRG.

Questions remain that new observations could answer.
High-resolution imaging could determine if the starburst is in a circumnuclear
ring, a second nucleus, or in a nearby galaxy in a chance alignment.
More spectroscopy could determine if the companion object is an interacting
galaxy or an intervening star.  

\section{Conclusions}

UN J1025$-$0040 is a galaxy playing host to both a quasar and a massive 
starburst.  There is intrinsic polarization $>$1\%, increasing toward long 
wavelengths, perhaps arising from transmission through dust associated with 
the starburst.  This spectacular object may represent a transitional step 
in an evolutionary sequence between ULIRGs and optically selected quasars.

\acknowledgments

We thank Hien Tran, Ed Moran, Arjun Dey and the Keck staff for their 
assistance, and Carlos De Breuck for explaining our discovery to reporters.  
Data presented herein were obtained at the W. M. Keck Observatory, which is 
operated as a scientific partnership among the California Institute of 
Technology, the University of California, and the National Aeronautics and 
Space Administration.  The Observatory was made possible by the generous 
financial support of the W.M. Keck Foundation.
This work has been performed under the auspices of the U.S. Department of Energy
by Lawrence Livermore National Laboratory under Contract W-7405-ENG-48.
A.V.F. acknowledges the support of NASA grant NAG 5-3556.
We thank the Anglo-Australian Observatory for
their support of the 2dF QSO Survey. S.M.C. thanks the UK Particle Physics
and Astronomy Research Council for support.

\begin{deluxetable}{lccc}
\tablecaption{Photometry of UN J1025$-$0040}
\tablewidth{0pt}
\tablehead{Property & Observed & ISB Model\tablenotemark{a} & QSO Residual\tablenotemark{b}}
\startdata
$B$\tablenotemark{c} &           19.3    & 21.3 & 19.5 \nl
$M_B$\tablenotemark{d} &  $-$24.6\,\,\,\,\,\, & $-$23.8\,\,\,\,\,\, & $-$21.9\,\,\,\,\, \nl
$R$\tablenotemark{c} &           18.3   & 18.7  & 19.5 \nl
$K_s$\tablenotemark{e} &           16.7   & 17.0  & 18.5 \nl
\enddata
\tablenotetext{a}{Instantaneous starburst (ISB) model as in Fig. 3.}
\tablenotetext{b}{The difference between the observations and the ISB model.}
\tablenotetext{c}{Based on the narrow-slit spectra, corrected for 
Galactic reddening ($E(B-V) = 0.06$ mag). $B$ = 19.1 (dereddened) based on 
our older photometry.}
\tablenotetext{d}{Using $H_0 = 75$ km s$^{-1}$ Mpc$^{-1}$ and $q_0$ = 0, 
rest-frame $B$.}
\tablenotetext{e}{Total, including the fuzz.}
\end{deluxetable}

\psfig{file=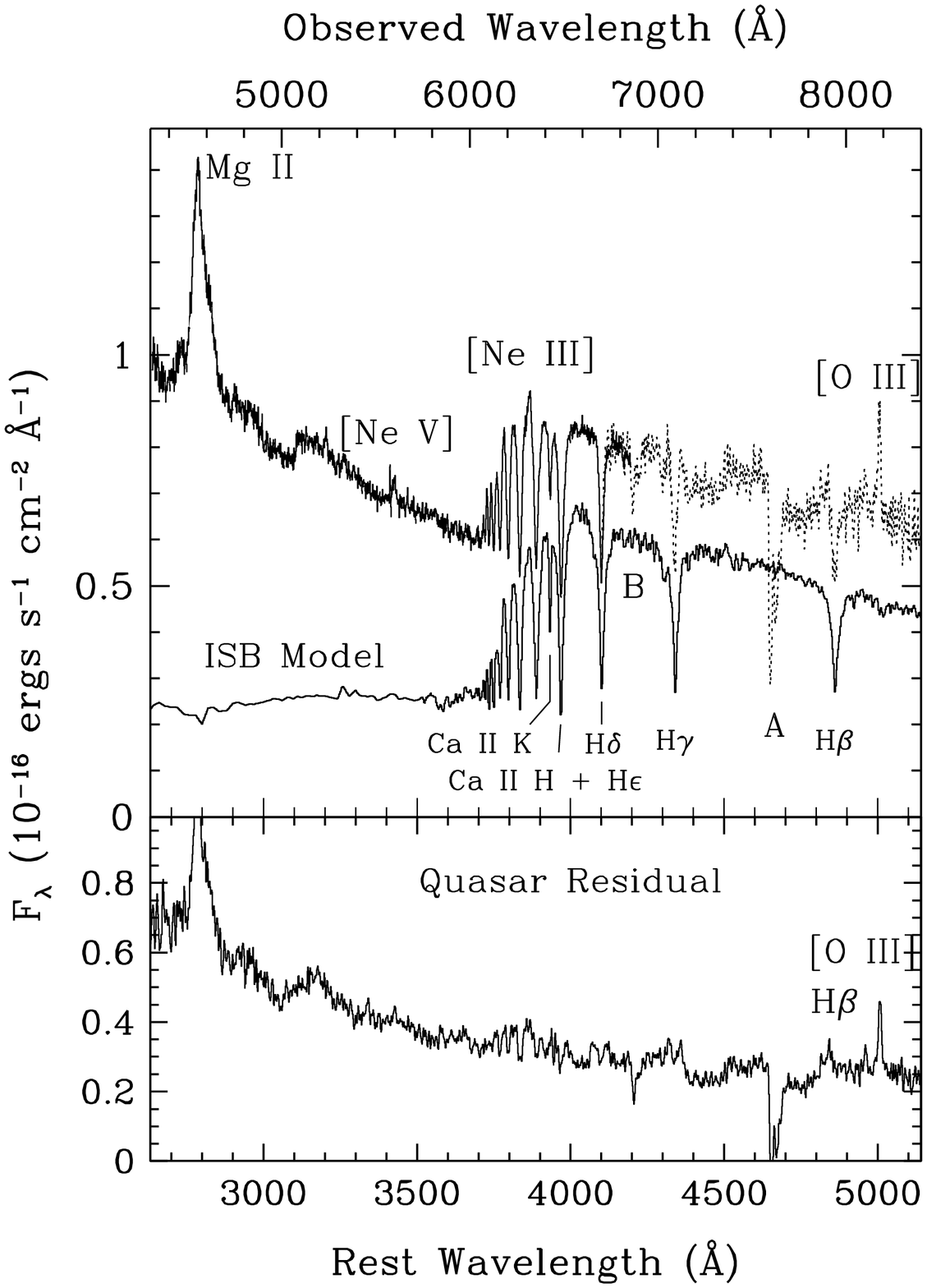,width=17cm}
\figcaption{
Keck spectropolarimetry total-light spectrum (60 minute exposure),
with the low-resolution discovery spectrum extending the wavelength coverage
(dotted line).  Rest-frame wavelengths assume $z=0.634$.  
Also plotted is a 400-Myr-old 
instantaneous starburst (ISB) model described in the text.
A and B mark telluric absorption.  The bottom panel shows the residual quasar
spectrum obtained by subtracting the ISB model from the low-resolution 
spectrum, revealing broad H$\beta$ emission.}

\psfig{file=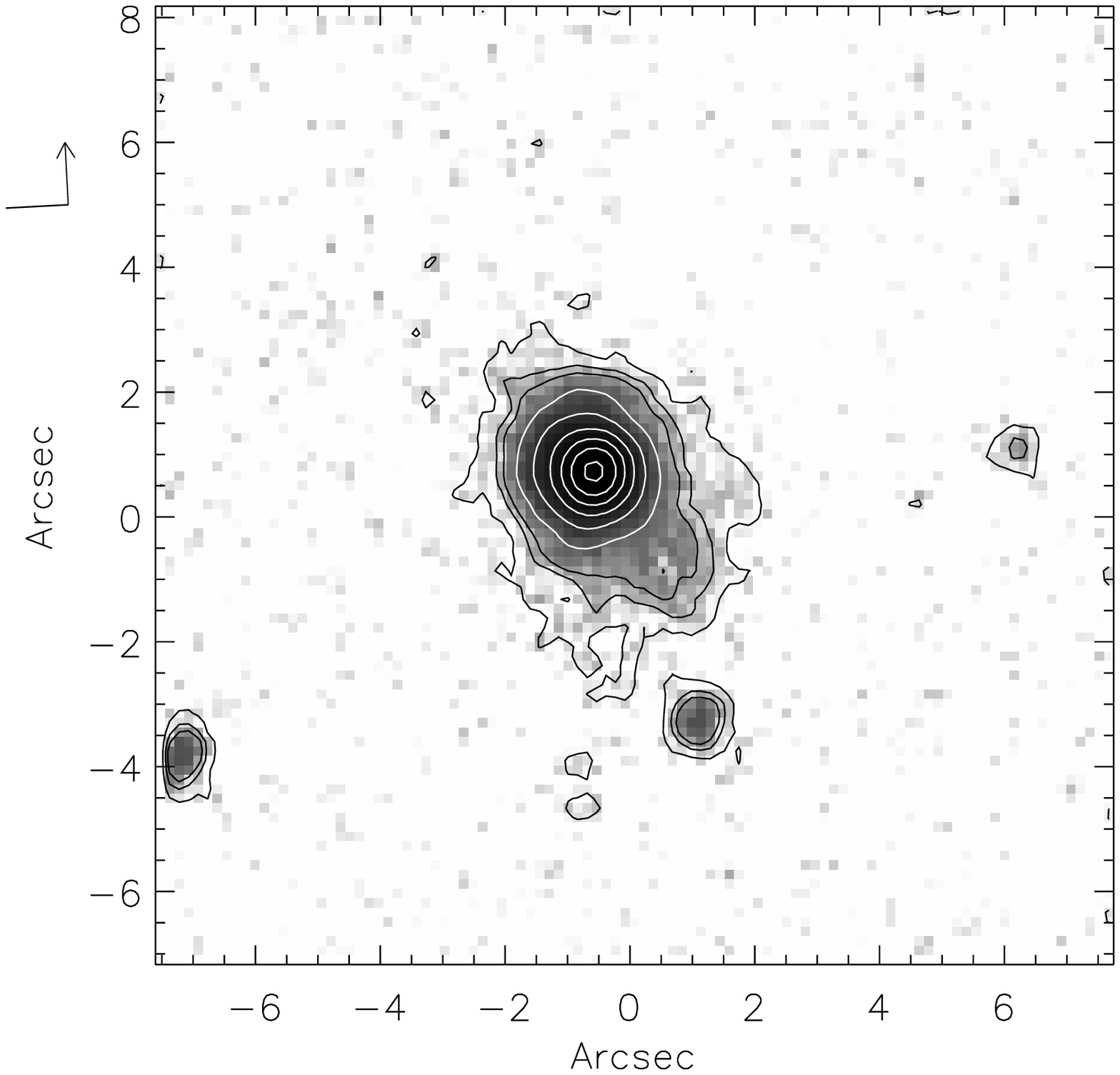,width=18cm}
\figcaption{
Keck $K_s$-band image
(32 min, $\sim0.5 \arcsec$ seeing). North (arrow) and East are indicated at
left.  The outer contour is 22.1 mag/square arcsec.
The scale is 4.6 $h^{-1}$\ kpc/arcsec ($q_o = 0$).
}

\psfig{file=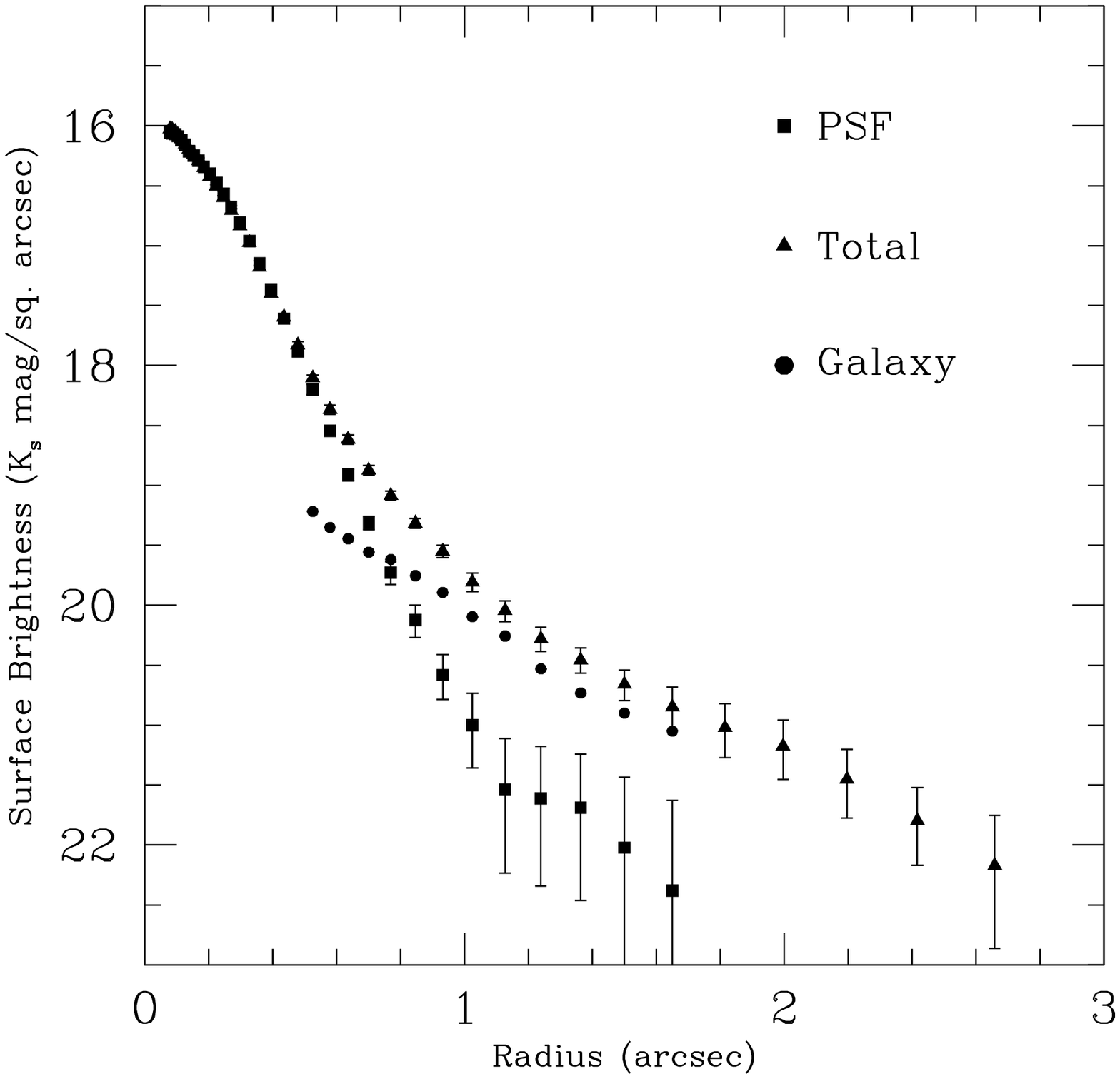,width=18cm}
\figcaption{
Surface brightness profile from ellipse
fitting, UN J1025$-$0040 (triangles) vs. a normalized point source (squares).
Filled circles show the difference between the total light and 70\% of
the normalized PSF.
}

\end{document}